\documentstyle[aps,preprint,epsf]{revtex}

\def\p{{\psi}}
\def\pb{{\overline\psi}}
\def\z{{\zeta}}
\def\zb{{\overline\zeta}}
\def\ga{{\gamma}}

\def\pdr{\partial}
\def\dsl{\not\partial}

\def\zbz{{\langle\zb\z\rangle}}

\begin{document}
\setcounter{page}{0}
\def\footnoterule{\kern-3pt \hrule width\hsize \kern3pt}
\tighten
\title{$Z_3$ TWISTED CHIRAL CONDENSATES IN QCD \\
	AT FINITE TEMPERATURES\thanks
{This work is supported in part by funds provided by the U.S.
Department of Energy (D.O.E.) under cooperative
research agreement \#DF-FC02-94ER40818.}}

\author{Shailesh Chandrasekharan
\footnote{Email address: {\tt chandras@mitlns.mit.edu}}
	and Suzhou Huang
\footnote{Email address: {\tt shuang@mitlns.mit.edu}}}

\address{Center for Theoretical Physics \\
Laboratory for Nuclear Science \\
and Department of Physics \\
Massachusetts Institute of Technology \\
Cambridge, Massachusetts 02139}

\date{MIT-CTP-2495,~ hep-ph/9512323. {~~~~~} December 1995}
\maketitle

\thispagestyle{empty}

\begin{abstract}
It was recently observed in a lattice QCD measurement that the
chiral condensate in the quenched approximation shows dramatically
different behavior in the three $Z_3$-equivalent de-confined phases.
We argue that this phenomenon can be understood qualitatively as an
effect of $Z_3$ twists on fermionic fields. Quarks under these $Z_3$-twists
become global anyons and hence display different thermodynamic properties.
We further show that the lattice data can be roughly modeled
by a Nambu-Jona-Lasinio type Lagrangian with a minimal coupling to
a constant gauge field $A_0=2\pi nT/3$ (with $n=0$,
$\pm1$), which arises naturally from the non-trivial phase of the
Polyakov line.

\end{abstract}

\vspace*{\fill}
\begin{center}
Submitted to: {\it Physics Letters B}
\end{center}

\pacs{xxxxxx}

\section{Introduction}
  Understanding chiral symmetry breaking and its restoration at finite
temperatures in QCD is of great interest. Many lattice simulations have
been performed to study the expected phase transitions
\cite{ref:1 Kanaya}.
However, the underlying dynamics responsible for these phenomena are still
unclear. Hence, it is sometimes useful to explore regions of QCD parameters
beyond their physical domain. One such effort in lattice QCD
is to study the fermionic Greens functions in the background dynamics of
pure gauge fields. In such a study one ignores all the effects of internal
quark loops. At low temperatures no essential difference is
observed between such studies and those that take into account the effects
of quark loops. However, at finite temperatures these effects can change
the physics qualitatively. In the pure gauge theory there is a
first order phase transition between a confined and a de-confined phase
signaled by a non-vanishing expectation value of the Polyakov line.
This is related to a global $Z_3$ symmetry breaking\cite{ref:2 Yaffe}.
On the other hand, inclusion of massless quarks breaks this global
$Z_3$ symmetry explicitly and chiral symmetry restoration becomes relevant,
with the chiral condensate as the order parameter.
The precise nature of the chiral transition
depends on the number of flavors\cite{ref:3 Pisarski} appearing through
internal quark loops.

Interestingly, in lattice QCD one can measure a {\em quenched chiral
condensate} even after ignoring internal quark loops from the dynamics.
Such a condensate has been measured in the past near the de-confining
transition. A first order transition was observed
in the {\em quenched chiral condensate} at the de-confining temperature
\cite{ref:4 Kogut}. However, no efforts were made to distinguish between
the behavior of quarks in the three de-confined phases
labeled by an element of $Z_3$. Recently, in an effort to understand the
effects of internal quark loops near the chiral phase transition, this issue
was revived by a
careful and precise measurement of the {\em quenched chiral condensate}
\cite{ref:5 Sch}. It was observed that, above the de-confining transition,
the chiral condensate crucially depends on the $Z_3$ phase in which
the gauge dynamics settles. Even the chiral symmetry restoration
appears to occur at different temperatures, depending on the phase. This
somewhat surprising
result needs an explanation. In this article we present a qualitative
understanding of the physics underlying these observations.

In a somewhat different context the effect of the boundary condition
in the Gross-Neveu model, with one of the dimension compactified, was
studied recently\cite{ref:6 Suzh}. It was found there that the fermion,
which obeys the usual
antiperiodic boundary condition, becomes a global anyon when a non-trivial
twist is introduced via a constant background gauge field. The anyon
interpretation is supported by an explicit decomposition of the fermion
propagator into a sum over winding numbers in the compactified direction.
As a consequence of this statistics transmutation the thermodynamic properties
of the Gross-Neveu model become sensitive to the value of the
twist. We suggest that a similar physical setting is realized in quenched
QCD in the various de-confined phases. The non-trivial phase of the Polyakov
line provides a constant background gauge field in the temporal direction and
turns quenched quarks into global anyons.

In section 2, we show that in QCD one can define a chiral condensate with
a $Z_3$ twist in general. The usual chiral condensate is the one where the
twist is zero. The other two are new quenched order parameters of chiral
symmetry breaking. We then argue that the chiral
condensates measured in \cite{ref:5 Sch} are related to these $Z_3$ twisted
condensates in the quenched limit. In section 3, we elucidate the meaning
of the $Z_3$ twisted chiral condensate by considering an effective action
which governs these quenched observables. In section 4, we show that
the qualitative features of the lattice data can be obtained by the
Nambu-Jona-Lasinio model in the large $N_c$ limit, minimally coupled to
the suggested constant gauge field.
In section 5, we present some observations and conclusions.

\section{$Z_3$ twisted chiral condensates}
\label{z3cond}

  The chiral condensate in QCD is the order parameter for chiral symmetry
breaking and is defined formally as
\begin{equation}
\langle\pb\p\rangle
\;=\; {1\over Z} \int [dA]\;
tr{1\over D + m_q} \;Det( D + m_q)\; \exp(-S_g[A])
\end{equation}
where $D(A)\;=\; \ga^\mu(\pdr_\mu-iA_\mu)$ is the Dirac operator  and
$m_q$ is the quark mass and $S_g$ is the gauge action. $Z$ is the partition
function. The above definition can be made more precise on the lattice.

We now introduce a chiral condensate with a $Z_3$ twist $\theta$, given by
\begin{equation}
\label{eq:zbz}
\langle\zb\z\rangle_{\theta}
\;=\; {1\over Z} \int [dA]\;tr{1\over D_\theta + m_\z}\; Det( D + m_q)\;
\exp(-S_g[A])
\end{equation}
where
\begin{equation}
\label{eq:dtheta}
D(A)_\theta \;=\; \ga^\mu(\pdr_\mu\;-\;iA_\mu) + i\theta\;T\;\ga^0
\end{equation}
with $\theta=0,\pm {2\pi\over 3}$ and $T$ the temperature. The mass in
the observable, $m_\z$, is deliberately chosen to be distinct from the usual
quark mass, $m_q$. Thus $\zbz$ is a quenched observable and can be measured
on the lattice even in the presence of dynamical fermions.
The physical content of this new observable will be elucidated shortly.
Notice that $D_\theta$ is simply the original Dirac operator with the
temporal gauge field shifted by a phase of $Z_3$. Further note that
$\langle\zb\z\rangle_\theta\;=\;\langle\zb\z\rangle_{-\theta}$. We can show
this by noting that $D(A)$ and $D(-A)$ have the same spectra due to a
charge conjugation symmetry. However, in general we will have
$\langle\zb\z\rangle_{2\pi\over 3}\;\neq\;\langle\zb\z\rangle_{0}$.
This is related to the fact that fermions break the $Z_3$ symmetry.

%Here it is natural to study the
%chiral condensate in the three different $Z_3$ phases. The chiral
%condensate in each of the $Z_3$ phase, labeled by $\theta$, is the same as
%$\lim_{m_q\rightarrow \infty}\;\;\zbz_{\theta}$. To see this note that

In general, the dynamical quarks in eq.~(\ref{eq:zbz}), will always pick
the phase in which the gauge field distribution, $\{A_\mu\}$, makes the
expectation value of the Polyakov line $P$ real and positive, i.e.
\begin{equation}
\langle P({\bf x})\rangle\;=\;\left\langle{1\over 3}
Tr\left(\;P\int_0^{1\over T} \exp( iA_0({\bf x},t)\;dt) \right)\right\rangle
\;=\; |\langle P \rangle|
\end{equation}
The gauge field distribution of a $Z_3$ rotated phase
would be $\{A_\mu\}\;+\;\theta\;T\;\delta_{\mu,0}$. Now the Polyakov
line is given by $|\langle P \rangle|\;e^{i\theta}$. The relevant Dirac
operator for measuring the chiral condensate is $D(A)_\theta$ defined
in eq.~\ref{eq:dtheta}. Thus measuring the usual chiral
condensate in the various $Z_3$ phases is
naturally equivalent to measuring $\zbz_{\theta}$ in the quenched
limit of the de-confined phase.

The effect of similar $\theta$ terms on the thermodynamics has been
studied in the past for simple local fermionic field theories
\cite{ref:6 Suzh}. There it is argued that the $\theta$ term acts like
a twist in the boundary condition for the fermions and converts them into
global anyons. This conversion then affects the critical temperature
dramatically. We suggest that the same physics also happens in the
present context. We anticipate that, if we study
$\langle\zb\z\rangle_\theta$ as a function of the temperature, it is
likely that the condensate with $\theta=0$ and $\theta=\pm{2\pi\over 3}$
will behave differently. In particular if there is a phase transition in
the $m_\z\rightarrow 0$ limit it is quite likely that it will occur at
different temperatures depending on $\theta$.

 The above discussion then suggests that it is likely that at some
temperature in the de-confined phase $\langle\zb\z\rangle_0\; = \;0$
whereas $\langle\zb\z\rangle_{{2\pi\over 3}}\;\neq\;0$, as observed in
\cite{ref:5 Sch}. However, in the confined phase the gauge dynamics
is $Z_3$ symmetric in the quenched limit. The $Z_3$ twist in this
phase ceases to be meaningful, since the gauge field ensemble is
no longer clustered around any of the three $\theta$ values there.
%However if this is due to a shift in the critical
%temperature, as a result of the $Z_3$ twist, it would suggest that
%$\langle\zb\z\rangle_0\;\neq 0$ below some critical temperature.
%Unfortunately in the case of QCD the de-confined phase is a finite
%temperature phase and ends at some non-zero temperature $T_c$. The phase
%below $T_c$ is confining and the above discussion of the de-confined phase
%fails. Here the gauge field dynamics is $Z_3$ symmetric and the different
%$Z_3$ twisted chiral condensates become equal.
It is of course still possible to study the $Z_3$ chiral condensates, as
defined in eq.~\ref{eq:zbz}, even in the presence a dynamical quark mass
$m_q$, because of the relative shift in the Dirac operator in the observable
and in the internal quark determinant. In
the presence of finite $m_q$ the $Z_3$ symmetry is broken and thus it is
possible to track the three phases into the confined phase and study their
evolution towards the chiral limit. We do not know if such a study is
physically relevant since the $Z_3$ condensates are in general some
quenched order parameters. In the remainder of this
article we will be interested only in the quenched limit.

\section{An Effective Action for the $Z_3$ Twisted Condensates}

  In this section we suggest a local field theory which governs the
behavior of the $Z_3$ twisted chiral condensates. Formally one can
rewrite eq.~(\ref{eq:zbz}), introducing fermionic ($\z$) and bosonic
($\phi$) ghost fields in the spirit of \cite{ref:7 Golt},
\begin{equation}
\langle\zb\z\rangle_{\theta}
\;=\; {1\over Z} \int_{\psi,\z,\phi,A}\;\zb\z\;
\exp\left[-S_g[A] + \pb\;(D+m_q)\p + \zb(D_\theta+m_\z)\z +
\phi^\dagger(D_\theta+m_\z)\phi \right]
\end{equation}
This enlarged theory contains another chiral symmetry in the limit
$m_\z\rightarrow 0$ for arbitrary real quark mass $m_q$. It is now
obvious that the order parameter associated with this symmetry is
$\langle\zb\z\rangle$. One can then formally integrate over the quark
fields $\psi$, the ghost fields $\phi$ and the gauge fields $A_\mu$
to generate an effective action for the $\z$ fields. Thus
\begin{equation}
\label{eq:effzbz}
\langle\zb\z\rangle_{\theta}
\;=\; {1\over Z} \int_{\z}\;\zb\z\;
\exp\left[-S^{eff}_{[m_q,\theta,m_\z,T]}(\zb,\z) +
\zb(\ga^\mu\pdr_\mu\;+i\theta\;T\;\ga^0+m_\z)\z \right]
\end{equation}
We have allowed for the possibility that the effective action could depend
on the temperature through the couplings apart from the usual anti-periodic
boundary conditions in the Euclidean time direction for the $\z$ fields.

In general $S^{eff}_{[m_q,\theta,m_\z,T]}(\zb,\z)$ would be arbitrarily
complicated and possibly contains non-local interactions. It seems
hopeless to understand the full structure of the effective action.
However, it may be interesting to obtain some qualitative insight into
the essential physics involved by studying a simplified model that is
motivated phenomenologically and at the same time is tractable.

 In the quenched limit, the chiral symmetry between fermionic Greens
functions is governed only by the dynamics of the gauge fields.
Non-perturbative gauge fields can in general
produce chiral symmetry breaking. One immediate consequence is that the
quenched chiral condensate need not vanish. This feature, in principle,
need not have any connection with confinement. Thus it is not unnatural
to assume that, for some temperature range, chiral symmetry breaking
happens even in the deconfined phase. Consequently we are led to study
chiral symmetry breaking in the absence of confinement. To this end,
we assume that the effective interaction generated after integrating
out all other fields is dominated by the lowest dimensional term,
\begin{equation}
\label{eq:njlint}
S^{eff}_{[m_q,\theta,m,T]}(\zb,\z) \;=\;
{G(T)\over 2N_c \Lambda^2}\;
\left[(\zb\cdot\z)^2 - (\zb\cdot\ga_5\z)^2\right]
\end{equation}
with $\Lambda$ being some cutoff scale specified later. Thus
in this model the $\theta$ dependence enters only through the quadratic
term shown in eq.~\ref{eq:effzbz}. We expect this term to produce the
difference between the complex and the real phases as in \cite{ref:6 Suzh}.

  As mentioned earlier, in the confined phase there is no difference between
the various $Z_3$ twisted condensates. This is due to the $Z_3$ symmetry of
the dynamics. Thus the effective action must have no knowledge of $\theta$.
This is in accordance with the fact that the expectation value of the Polyakov
line is zero in the confined phase, it is meaningless to include a $\theta$
term in the confined phase. Here no attempt is made to construct an effective
field theory that describes the confinement physics also in a natural way.
However, as long as we are interested only in a model for the chiral
condensate,
it may be sufficient to drop the $\theta$ term from the effective action in
the confined phase. Additional effects of gauge dynamics are mimicked by a
temperature dependence of the coupling $G(T)$. Further since the de-confining
transition is a first order transition, an abrupt jump is possible in $G(T)$
at the de-confinement temperature.

\section{Modeling of lattice data}
  To have a semi-quantitative understanding of the lattice data
we start with the Lagrangian
\begin{equation}
\label{eq:njl}
{\cal L}(\zb,\z)\;=\; \zb\cdot\dsl\z\;+\;
\zb\cdot\ga^0\z \; i\theta\;T\;+m\;\zb\z \;+
{G(T)\over 2N_c \Lambda^2}\;\left[(\zb\cdot\z)^2
-(\zb\cdot\ga_5\z)^2\right]
\end{equation}
or equivalently
\begin{equation}
{\cal L}(\zb,\z)\;=\; \zb\cdot\dsl\z\;+\;
\zb\cdot\ga^0\z \; i\theta\;T\;+m\;\zb\z \;+
\zb(\sigma+i\gamma_5\pi)\z\;+{N_c\Lambda^2\over 2G(T)}(\sigma^2+\pi^2)
\end{equation}

Since this model is non-renormalizable we consider a momentum cutoff
in the spatial momentum $|{\bf p}|=\Lambda$. We will solve the model in
the large $N_c$ limit and fix the parameters in such a way that the
model reproduces the lattice data obtained in \cite{ref:5 Sch}
qualitatively. The effective potential in the large $N_c$ limit is given by
\begin{equation}
V_{\text{eff}}(\sigma)={N_c\Lambda^2\over 2G(T)}\sigma^2
-2N_c\int^\Lambda{d^3 p\over(2\pi)^3}\; T\sum_{n=-\infty}^\infty
\ln\left[\tilde\omega_n^2+p^2+(\sigma+m)^2\right]
\end{equation}
where
\begin{equation}
\tilde\omega_n=[(2n-1)\pi+\theta]T
\end{equation}
The Matsubara sum can be carried out by the standard contour
integral technique, yielding
\begin{equation}
\label{eq:dveff}
{\partial\over\partial\sigma}V_{\text{eff}}(\sigma)=
{N_c\Lambda^2\over G(T)}\;\left\{\sigma-{G(T)\over \pi^2\Lambda^2}
(\sigma+m)\;\int^\Lambda_0 dp\; {p^2\over E_p}\left[1-
{1\over e^{E_p/T+i\theta}+1}-{1\over e^{E_p/T-i\theta}+1}\right]\right\}
\end{equation}
with $E_p=\sqrt{p^2+(\sigma+m)^2}$.
In the above equation the anyon behavior is clearly displayed from
the modified thermo-distribution, which interpolates smoothly from the
Fermi-Dirac ($\theta=0$) to Bose-Einstein ($\theta=\pi$) distributions.

The gap equation in the large $N_c$ limit reduces to
\begin{equation}
{\sigma\over G(T)}\;=\;
{\sigma+m\over\pi^2\Lambda^2}\int_0^{\Lambda} {p^2 dp\over E_p}\;
\left[1-{1\over e^{E_p/T+i\theta}+1}-{1\over e^{E_p/T-i\theta}+1}\right]
\end{equation}
The full effective potential can be obtained after integrating
eq.~\ref{eq:dveff},
\begin{equation}
V_{\text{eff}}(\sigma)-V_{\text{eff}}(0)
={N_c\Lambda^2\over G(T)}\left\{{\sigma^2\over 2}
-{G(T)T\over \pi^2\Lambda^2}\;\int_0^\Lambda p^2\, dp\;
\ln\left[{\cosh(\sqrt{p^2+(\sigma+m)^2}/T)+\cos\theta \over
\cosh(\sqrt{p^2+m^2}/T)+\cos\theta}\right]\right\}
\end{equation}
We refer the reader to
\cite{ref:6 Suzh} for a detailed discussion of the effects of $\theta$.
Here we only try to model the lattice data of \cite{ref:5 Sch}.

  The parameters in the NJL model are fixed in a conventional way
\cite{ref:75 hatsuda}. Assuming the cutoff $\Lambda$ to be about 1 GeV,
we set $\Lambda/T_c=5.5$ and $G(T)/(2\pi^2)=1.163$ for $T<T_c$ in order
to reproduce the chiral condensate at low temperatures with $T_c$ around
200 MeV. The $\theta$ term is dropped in the confined phase for reasons
already explained. The current quark mass
on the lattice $(m_\z)$, and in the effective action $(m)$ can in
general be different.
Here we assume\footnote{In addition to contributions proportional to
$m_\z$, $m$ could get other contributions, such as from the $U_A(1)$
anomalies. We will ignore such additive contributions here.}
$m\;=\;Z\;m_\z$. Then $Z=1.48$ can be cleanly determined by fitting
the condensate ($\theta=0$) in the chirally symmetric phase.
However, in order to reproduce the lattice data quantitatively
a non-trivial $T$-dependence in $G(T)$ needs to be introduced
in the deconfined phase. The first order nature of the de-confining
transition is incorporated by a jump in $G(T)$. Then a simple quadratic
form $G(T)/(2\pi)^2=1.385-0.385(T/T_c)^2$ is sufficient for $T>T_c$.
\begin{figure}[b]
\epsfxsize=130mm
\epsffile{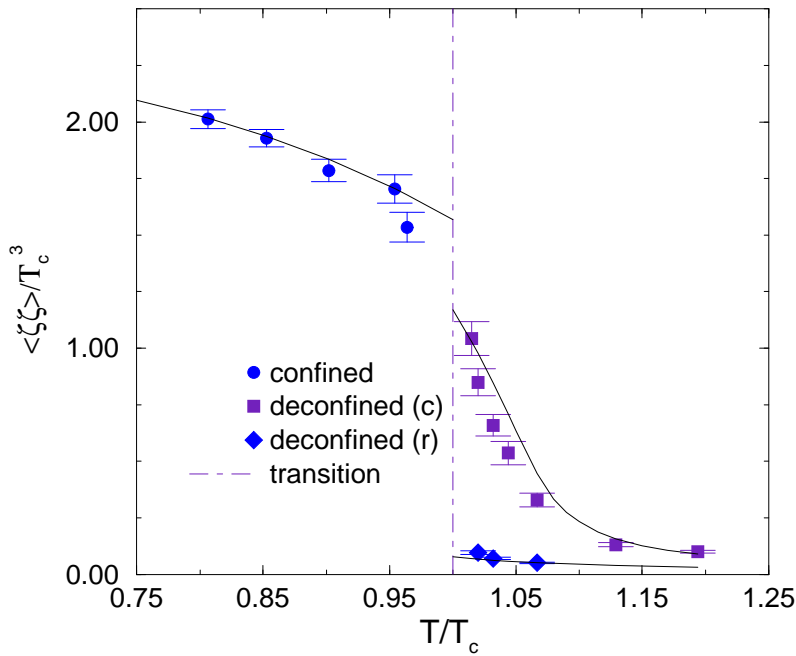}
\caption{
This graph shows the lattice data. The behavior of the condensate
in the large $N_c$ Nambu-Jona-Lasinio model described in the text,
produces the solid lines. The lattice data corresponds to
$m_\z\;a\;=\;0.001$, where $a\;=\;{1\over T\;N_t}$ with $N_t\;=\;4$.
\label{fig:qphase1} }
\end{figure}

The lattice data corresponds to $N_t=4$ and the lattice
spacing is given by $a\;=\;{1\over T\;N_t}$. Then the value of the
condensate computed from the NJL model should be compared with
the lattice data as
\begin{equation}
-{1\over N_c}\langle\zb\cdot\z\rangle\;
	=\;{\sigma\Lambda^2\over G}\;=\;
{1\over G}{\sigma\over\Lambda}\left({\Lambda\over T_c}\right)^3\;T_c^3
\;=\;{\sigma\over\Lambda}\;{5.5^3\over G}\; T_c^3
\end{equation}
where $\sigma$ is obtained from the solution of the gap equation.

\begin{figure}[hbt]
\epsfxsize=130mm
\epsffile{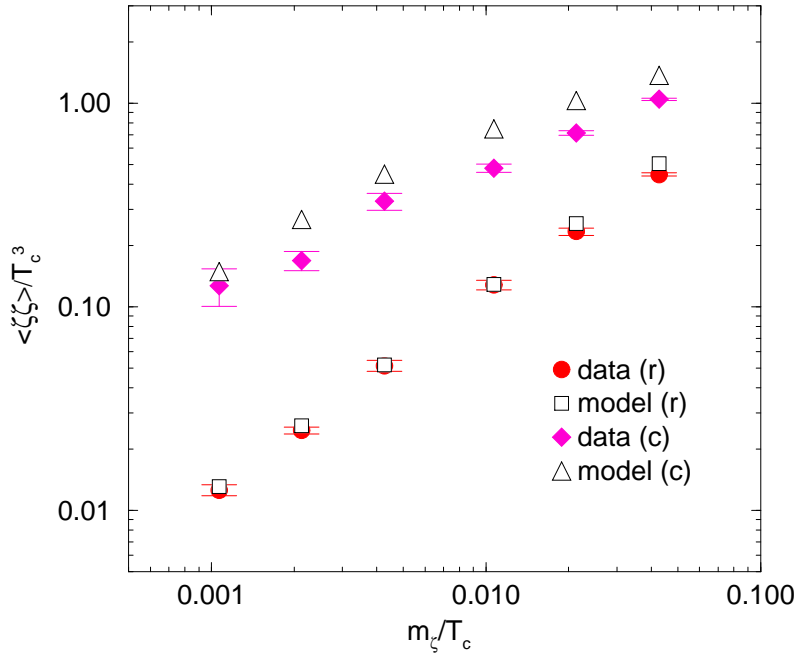}
\caption{
This graph compares the behavior of the condensate as a
function of the current quark mass at $T=1.067T_c$ in the large $N_c$
Nambu-Jona-Lasinio model described in the text with the lattice data.
This is approximately the temperature where a possible transition occurs
in the complex phase. (r) refers to $\theta=0$ and (c) refers to
$\theta=\pm 2\pi/3$.
\label{fig:qphase2} }
\end{figure}

The data
from lattice simulations\cite{ref:8 Sch} and the model are given in
figures~\ref{fig:qphase1} and \ref{fig:qphase2}.
\footnote{ The actual data was given in
terms of the lattice coupling. This has been converted to dimensionless
less units using the critical temperature as the scale. The bare lattice
coupling $\beta=5.692$ was used as the critical coupling for the de-confinement
temperature.} It is seen there that the NJL model, with its parameters
properly tuned, is capable of describing the data semi-quantitatively.
However, we would like to stress the fact that the final outcome of the
NJL model sensitively depends on the precise temperature dependence. A
small change in the numerical value of the coupling $G$ can change the
evolution of the complex chiral condensate drastically. Thus in this regard
we find the model somewhat unnatural, even though it is possible to capture
the qualitative physics.

In figure~\ref{fig:qphase2} the current mass dependence of the condensate
is compared between the model and the lattice data at $T/T_c\;=\;1.067$.
At this temperature the complex phase lattice data begins to show interesting
power law dependence, and the real phase lattice data is essentially linear
\cite{ref:8 Sch}. The real phase data at $T=1.067\;T_c$ was used to fix the
constant $Z$ and hence will agree well with the model. However, it is obvious
that the present model will not reproduce the power law behavior of the real
phase lattice data at lower temperatures\cite{ref:5 Sch}. Similarly we do
not expect the model to
reproduce the power law behavior of the complex phase either. In fact we do
not know if any universality arguments apply to these data. However, we do
see the qualitative difference between the real and the complex phases.

\section{Discussion and Conclusions}

  In this work we have clarified the precise meaning of the chiral
order parameter considered in a recent lattice study. The observed
$Z_3$ phase dependence of this order parameter in quenched QCD
in the deconfined phase has been qualitatively explained.
The relevant physical picture is that the non-trivial $Z_3$ twists
in the temporal boundary condition, due to the survival of expectation
value of the Polyakov line, turn quarks into global anyons. The
qualitative behavior of the lattice data can be modeled simply by
a temperature dependent NJL model with a minimal coupling to a
constant gauge field $A_0=\theta\,T$ with $\theta=0$ and $\pm 2\pi/3$.

  For $m\;=\;0$, the NJL model would predict a
continuous phase transition when $\theta=\pm 2\pi/3$,
in addition to the jump associated with the first order de-confining
phase transition. The present lattice data are consistent with
this prediction. The second transition temperature is estimated
to be about 5 to 10 percent higher than the de-confining transition
temperature.

  As this work was being completed we received an article\cite{ref:9 Ogil},
which discusses the same issue along very similar lines using the NJL
model. However, based on our calculation, we would like to point out that,
on a quantitative level the model does not reflect one important feature
of the gauge dynamics. It is clear from figure~\ref{fig:qphase1} that the
lattice data for the complex phase evolves rapidly above $T_c$ towards a
possible second transition at about $T=1.07(3)\;T_c$. In the above model
such a rapid evolution required very fine tuning of the coupling $G(T)$.
Thus it is very likely that the gauge dynamics is not {\em naturally}
captured in this model. In fact the rapid evolution just mentioned is also
seen in other physical quantities like the entropy density\cite{ref:10 Brow}
in the same region. Thus an obvious direction to extend the present
work, is to obtain a better model including the gauge dynamics. This may
help in understanding the reason for the rapid evolution seen just above
$T_c$.

  As was suggested in section~\ref{z3cond}, the $Z_3$ condensates as
defined by eq.~\ref{eq:zbz} can be studied even in the presence of small
dynamical quark masses $m_q$. As quenched order parameters these may be
measured in lattice simulations. However, more theoretical work would be
necessary to see if they are interesting. Finally, in this article we
have concentrated on the lattice data for $SU(3)$ gauge theories. A similar
discussion in the case of $SU(2)$ leads to interesting possibilities since
the de-confining transition is second order in this case. Extensions to
$SU(N)$ with $N > 3$ may also be interesting and has been attempted in
\cite{ref:9 Ogil}.

\section{Acknowledgments}
We would like to thank the Columbia Lattice group for allowing us to
use the lattice data prior to their publication. These have been presented
at the recent lattice conference\cite{ref:5 Sch} and the detailed results
will be published shortly \cite{ref:8 Sch}. We would
also like to thank U. -J. Wiese for helpful discussions.

\end{document}